\documentclass[pra,showpacs,floatfix,superscriptaddress,twocolumn]{revtex4}
\usepackage{graphicx}
\usepackage{epstopdf}
\usepackage{amssymb}
\usepackage{amsmath}
\usepackage{dsfont}
\usepackage{xspace}
\usepackage{braket}
\usepackage{color}
\usepackage{soul}
\usepackage{xcolor}
\usepackage{CJK}

\renewcommand{\k}[1]{\vert #1\rangle}
\renewcommand{\b}[1]{\langle #1\vert}
\newcommand{\up}{\ket{\uparrow}}
\newcommand{\down}{\ket{\downarrow}}
\newcommand{\carb}{$^{13}\mbox{C}$~}
\newcommand{\NVm}{NV$^-$}

\newcommand{\nitro}{$^{15}\mbox{N}$}

\begin{document}

\title{Probabilistic magnetometry with two-spin system in diamond}
\author{Ra\'ul Coto}
\email{raul.coto@umayor.cl}
\affiliation{Centro de Investigaci\'on DAiTA Lab, Facultad de Estudios Interdisciplinarios, Universidad Mayor, Chile}
\author{Hossein T. Dinani}
\affiliation{Centro de Investigaci\'on DAiTA Lab, Facultad de Estudios Interdisciplinarios, Universidad Mayor, Chile}
\affiliation{Instituto de F\'{\i}sica, Pontificia Universidad Cat\'{o}lica de Chile,
Casilla 306, Santiago, Chile}
\author{Ariel Norambuena}
\affiliation{Centro de Investigaci\'on DAiTA Lab, Facultad de Estudios Interdisciplinarios, Universidad Mayor, Chile}
\author{Mo Chen}
\affiliation{
Research Laboratory of Electronics, Massachusetts Institute of Technology, Cambridge, Massachusetts 02139, USA
}
\affiliation{
Department of Mechanical Engineering, Massachusetts Institute of Technology, Cambridge, Massachusetts 02139, USA
}
\author{Jer\'onimo R. Maze}
\affiliation{Instituto de F\'{\i}sica, Pontificia Universidad Cat\'{o}lica de Chile,
Casilla 306, Santiago, Chile}
%\affiliation{Research Center for Nanotechnology and Advanced Materials CIEN-UC, Pontificia Universidad Cat\'olica de Chile, Santiago 7820436, Chile}

\begin{abstract}
Solid-state magnetometers like the Nitrogen-Vacancy center in diamond have been of paramount importance for the development of quantum sensing with nanoscale spatial resolution. The underlying protocol is a Ramsey sequence, that imprints an external static magnetic field into the phase of the quantum sensor, which is subsequently read out. In this theoretical work we propose a sensing scheme that harnesses the hyperfine coupling between the Nitrogen-Vacancy center and a nearby nuclear spin to set a post-selection protocol. We show that concentrating valuable sensing information into a single successful measurement yields an improvement in sensitivity over Ramsey in the range of short transverse relaxation times. By considering realistic experimental conditions, we found that the detection of weak magnetic fields in the $\mu$T range can be achieved with a sensitivity of few tens of nTHz$^{-1/2}$ at cryogenic temperature ($4$ K), and $\mu$THz$^{-1/2}$ at room temperature. 
\end{abstract}

\maketitle

\section{Introduction}

Quantum metrology takes advantage of the quantum properties of the system to achieve better precisions than allowed by its classical counterpart~\cite{Escher11}. In particular, probabilistic quantum metrology aims to further increase the retrievable information of a parameter by means of a selective measurement. Inspired by the findings of Aharonov, Albert and Vaidman~\cite{Aharonov88}, where a measurement sequence consisting of pre-selection, weak measurement and  post-selection leads to an anomalous amplification of the measurement result, several theoretical works~\cite{Dressel14,Torres16,Coto17,Magana16,Arvidsson-Shukur19,Wu09,Brunner10,Feizpour11,Zhang15,Vitalie} and experiments~\cite{Ritchie91,Pryde05,Dixon09,Starling09,Kocsis11,Hosten08,Blok14,Alves17} have followed to explore such anomalous amplification. Nevertheless, there remains a longstanding controversy regarding whether probabilistic quantum metrology has practical advantages over standard techniques for parameter 
estimation~\cite{Combes14,Knee13,Knee14,Ferrie14,Zhang15,Tanaka13,Alves15}. In this work, we investigate the important case of spin magnetometry with color centers in diamond, and provide experimental parameter regimes where probabilistic quantum metrology is expected to succeed.

Quantum sensors, in particular solid-state magnetometers like the Nitrogen-Vacancy (NV) center in diamond have attracted widespread attention as a powerful tool at the nanoscale \cite{Schloss18,Arai18,Degen17,Taylor08,Maze08,Zaiser16,Aiello13,Balasubramanian09,Hirose12,Bonato}. Development of sensing protocols and experimental techniques facilitates detection of weak magnetic fields, featuring applications including sensing of single protein ~\cite{Lovchinsky16}, small molecules ~\cite{Glenn18}, single spins ~\cite{Sushkov14l,Shi15}, and more recently 3D reconstruction of a nuclear spin cluster ~\cite{Abobeih19}. 

In the following, we focus on DC magnetometry using a single NV center. The conventional measurement is realized by a Ramsey sequence with only the electronic spin of the NV. Here, we consider in addition a coupled nuclear spin located nearby the NV center. We show that by following a particular sequence involving post-selection one can achieve magnetic field sensitivity, the minimum detectable magnetic field normalized by the total sequence time, that is comparable with Ramsey sensitivity and outperforms the latter in suboptimal conditions. 

\section{Theory}

The original idea proposed by Aharonov, Albert and Vaidman \cite{Aharonov88} contributed to the field of probabilistic metrology, promoting the development of protocols that amplify the effect of the system-meter coupling strength to further estimate this coupling constant \cite{Dressel14,Magana16,Alves15,Alves17}. Moreover, some of these ideas have been implemented in color centers in diamond for estimation of the hyperfine coupling~\cite{Shikano} and weak measurements~\cite{Blok14,Pfender}. 

In this work, we propose a protocol consisting of pre-selection, system-meter interaction and post-selection. However, instead of amplifying the system-meter coupling strength itself, we take advantage of this interaction and use it to enhance the sensitivity for sensing an external magnetic field. We remark that a weak interaction is not required. We look for retrieving more information in a single successful estimation of the accumulated phase during the system-meter interaction. For convenience, we represent the initial state preparation and final state post-selection by three unitary rotations. The composite system then evolves according to $U=R_1(\theta_f)U_{\tau}R_1(\theta_i)R_2(\alpha)$, where the operator $R_1(\phi) (R_2(\phi))$ represents rotations of the system (meter) for an angle $\phi$ and $U_{\tau}$ is the free evolution of the system for time $\tau$ under the external magnetic field to sense. Thus, after post-selection upon the system in a target state $\ket{\psi_f}$, the meter state will be given by $\rho_{post}=\b{\psi_f}U\rho(0) U^{\dagger}\k{\psi_f}$, and the expectation value of the meter observable is
\begin{equation}
	\langle\sigma_2^z\rangle =\frac{\mbox{Tr}_2\left[\rho_{post}\sigma_2^z\right]}{\mbox{Tr}_2\left[\rho_{post}\right]},
\end{equation}
with $\sigma_2^z$ the Pauli $z$ operator acting on the meter. 

As a first step, we are interested in the shape of the signal $\langle\sigma_2^z\rangle $ that can be tuned by parameters $\lbrace \alpha, \theta_i, \theta_f \rbrace$ and the free evolution time $\tau$. This particular feature will be our starting point to enhance the magnetic field measurement, since it allows us to set the optimal interrogation time.

In the next section, we will show how this protocol can be implemented efficiently with a two-spin system in diamond at cryogenic temperature. The low temperature allows us to perform single-shot readout, which improves the signal-to-noise ratio. We will also discuss the scenario where the protocol is performed at room temperature.

\section{The Model}
\begin{figure}[ht]
\centering
\includegraphics[scale=0.45]{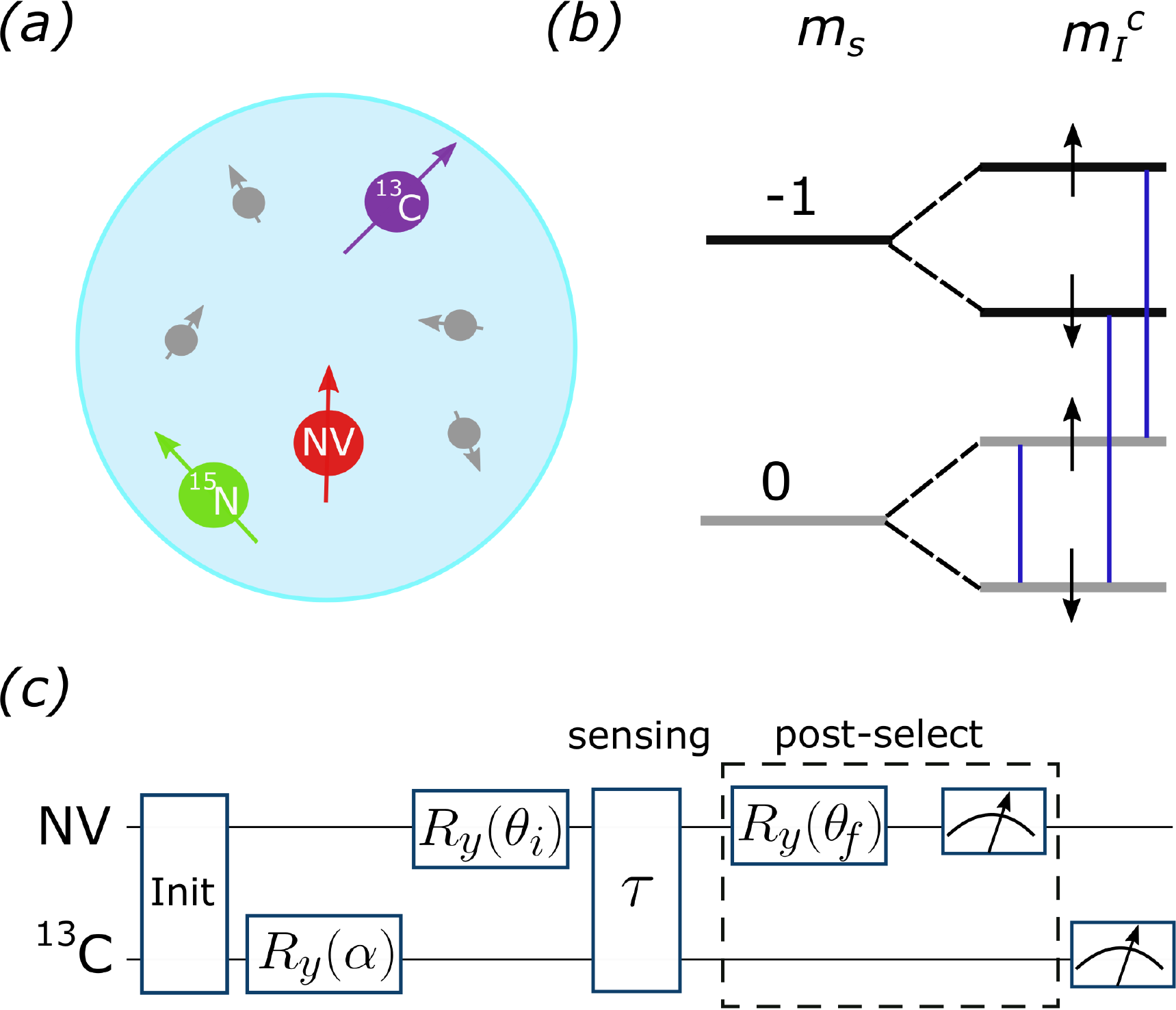}
\caption{(a) The electronic spin of a negatively charged NV center interacts with a nuclear spin corresponding to a Carbon-$13$. (b) Energy levels and relevant transitions. (c) Sequence for both electronic and nuclear spins.}
\label{fig1}
\end{figure}

For a more versatile sensing protocol that works both under room temperature and cryogenic temperature, we consider without loss of generality a concrete bipartite system model given by an electronic spin-$1$ ($S=1$) of a negatively charged Nitrogen-Vacancy center (NV$^-$) and a nearby nuclear spin-$1/2$ ($I=1/2$) of a Carbon-$13$ (\carb), as illustrated in Fig.~\ref{fig1} (a). Note that we have a native nuclear spin-$1/2$ (spin-$1$) from $^{15}$N ($^{14}$N) as well, which enables single-shot readout of the NV$^-$ for the post-selection at room temperature.

The ground state of the NV$^-$ is a spin triplet labelled by the spin quantum number $m_s = 0, \pm 1$ ($S = 1$). An external magnetic field $B_z$ along the N-V axis ($z$-axis) induces Zeeman energy splitting between the spin sublevels $m_s=+1$ and $m_s=-1$ and lifts their degeneracy. The \carb is hyperfine coupled to the \NVm center, yielding the system Hamiltonian ($\hbar$=1)
\begin{equation}\label{system_H}
H_0 = D S_z^2 + \gamma_e S_z (B_z + B) +  \gamma_c I_z (B_z + B) +  S_z A_{zz} I_z,
\end{equation}
where $D/2\pi = 2.87$ GHz is the zero-field splitting of the NV$^{-}$ center, $\gamma_e/2\pi \approx 2.8$ MHz/G,  and $\gamma_c/2\pi \approx 1.07$ kHz/G are the gyromagnetic ratios of the electron and \carb nuclear spins, respectively. $B$ is the small magnetic field to detect and $A_{zz}$ is the hyperfine coupling strength. We choose a weakly coupled \carb nuclear spin aligned close to the N-V axis, such that its anisotropic hyperfine coupling terms such as $A_{zx}$ are small and thus neglected here.

To further simplify the analysis, we focus on the 2-level submanifold $\lbrace \k{0} (m_s=0),\k{1} (m_s=-1) \rbrace$ for the electronic spin of the NV$^{-}$, while for the \carb we consider the complete basis $\up$ ($m_I^c=+1/2$) and $\down$ $(m_I^c=-1/2)$. In Fig.~\ref{fig1} (b) we show the energy levels of our configuration, indicating the relevant transitions  with blue solid lines. 

In order to manipulate the energy sublevels depicted in Fig.~\ref{fig1} (b), we apply a series of microwave (MW) and radiofrequency (RF) pulse sequences (square pulses) that result in the total Hamiltonian in a multi-rotating frame 

\begin{equation}\label{h_rotating}
\tilde{H}=
\frac{1}{2}\begin{pmatrix}
\gamma_c B & \Omega_{0}^{c} & \Omega^{e}& 0  \\
 \Omega_{0}^c & -\gamma_c B & 0 & \Omega^{e}  \\
 \Omega^{e}& 0 & 2\delta_{1}^{\uparrow}& 0  \\
0 & \Omega^{e} & 0  & 2\delta_{1}^{\downarrow}
 \end{pmatrix},
\end{equation}
where $\Omega^e$ and $\Omega^c_{0}$  are the Rabi frequencies of the MW and RF fields acting on the electron and nuclear spins, respectively. $\delta_1^\uparrow=-\gamma_eB-A_{zz}/2 +\gamma_c B/2$ and $\delta_1^\downarrow=-\gamma_eB+A_{zz}/2 -\gamma_c B/2$. For more details of the Hamiltonian and the rotating frame see Appendix \ref{Appendix_Rota}. 

In what follows, we describe our protocol that is represented in Fig.~\ref{fig1} (c), taking the electronic spin of the \NVm as the system and the nuclear spin as the meter. 
First, the bipartite system is initialized to the state $\k{\Psi_i}=\k{0}\otimes \ket{\downarrow}$. Efficient nuclear spin initialization has been demonstrated in Refs.~\cite{Taminiau14,Jamonneau16,Coto17b,Yun19,Bradley19}. Second, we prepare the nuclear spin in a coherent superposition state $\k{0}\otimes(\cos(\alpha/2)\up + \sin(\alpha/2)\down)$ via the RF field ($\Omega_0^c$). Third, a strong MW pulse rotates the \NVm electronic spin by an angle $\theta_i$ which is independent of the \carb nuclear spin state, yielding

\begin{eqnarray}\label{pre_selected}
\k{\Psi_{pre}} &=& (\cos(\theta_i/2)\k{1} + \sin(\theta_i/2)\k{0})\nonumber \\&\otimes&(\cos(\alpha/2)\up + \sin(\alpha/2)\down).
\end{eqnarray}

We will refer to Eq.~(\ref{pre_selected}) as the pre-selected state and it serves two goals in the sensing protocol: firstly, it acquires a phase proportional to $B$, directly contributing to the magnetometry. Additionally, it enables the interaction between the nuclear spin and the \NVm ~that is fundamental for probabilistic quantum metrology and the enhancement of the sensitivity. Next step, we let the system evolve for an interrogation time $\tau$, leading to

\begin{eqnarray}\label{state_evol}
\k{\Psi_1}&=& \cos(\theta_i/2)\cos(\alpha/2)e^{-i\delta_1^\uparrow\tau}\k{1}\otimes\up \nonumber\\ &+& \cos(\theta_i/2)\sin(\alpha/2)e^{-i\delta_1^\downarrow\tau}\k{1}\otimes\down \nonumber \\ &+& \sin(\theta_i/2)\cos(\alpha/2)e^{-i\gamma_cB\tau/2}\k{0}\otimes\up \nonumber\\ &+& \sin(\theta_i/2)\sin(\alpha/2)e^{i\gamma_cB\tau/2}\k{0}\otimes\down.
\end{eqnarray}

Finally the system is post-selected upon NV$^-$ in a target state $\k{\psi_f}=\cos(\theta_f/2)\k{1} + \sin(\theta_f/2) \k{0} $. This process is illustrated in Fig.\ref{fig1}-(c). In sensing a weak magnetic field $B$, we have $\gamma_c B\tau\ll 1$. Therefore the post-selection leaves the nuclear spin in the state 

\begin{eqnarray}\label{post}
\k{\bar{\phi}_{post}} &=& \cos(\alpha/2)\cos(\theta_f/2)\cos(\theta_i/2)e^{-i\delta_1^\uparrow\tau}\up \nonumber\\ &+&  \cos(\alpha/2)\sin(\theta_f/2)\sin(\theta_i/2)\up \nonumber\\
&+& \sin(\alpha/2)\cos(\theta_f/2)\cos(\theta_i/2)e^{-i\delta_1^\downarrow\tau}\down \nonumber \\ &+& \sin(\alpha/2)\sin(\theta_f/2)\sin(\theta_i/2)\down.
\end{eqnarray}
The above state needs to be normalized, such that $\k{\phi_{post}}=\k{\bar{\phi}_{post}}/\sqrt{P_s}$, where $P_s$ is the probability of having a successful post-selection,
\begin{eqnarray}\label{norm}
P_s &=& \frac{1}{2}\left[1+\cos(\theta_f)\cos(\theta_i)\right] + \frac{\sin(\theta_f)\sin(\theta_i)}{2}\nonumber \\&\times&\left[\cos^2 (\alpha/2) \cos(\delta_{1}^\uparrow\tau)+\sin^2(\alpha/2)\cos(\delta_1^\downarrow\tau)\right].
\end{eqnarray}

The final signal is proportional to
\begin{eqnarray}\label{exp_value}
\langle I_z\rangle &=& \frac{1}{4P_s} \left[1+\cos(\theta_f)\cos(\theta_i)\right] \cos(\alpha) \nonumber \\
&+& \frac{\sin(\theta_f)\sin(\theta_i)}{4P_s}\\ &\times&\left[ \cos^2 (\alpha/2) \cos(\delta_1^\uparrow\tau) - \sin^2(\alpha/2)\cos(\delta_1^\downarrow \tau)\right]\nonumber.
\end{eqnarray} 

We now compare the signal obtained in Eq.~(\ref{exp_value}) to the simple case of Ramsey spectroscopy $(\pi/2)_x-\tau-(\pi/2)_x$ considering a single spin, the \NVm electronic spin. The Ramsey signal follows $\langle S_z\rangle_R = 1/2-\cos(\gamma_eB\tau)/2$. Notice that our protocol shares a common ground with the Ramsey technique when sensing DC magnetic field as phase estimation. Furthermore, when the nuclear spin related part is removed, our protocol converges to the conventional Ramsey sequence with $\theta_i=\theta_f=\pi/2$. Nevertheless, our protocol employs a more elaborate procedure involving the additional nuclear spin, allowing further gain in standard deviation via the post-selection process at shorter times (Fig.~\ref{fig2} (b)). To emphasize the role of post-selection, we analyze the case where no post-selection is carried out. 
Since $\langle I_{z} \rangle=\cos(\alpha)/2$ carries no information about the magnetic field in this case, we calculate the expectation value of $I_x$ instead:
\begin{equation}
\langle I_x\rangle=\sin(\alpha)\left( \cos^2(\theta_i/2)\cos((A_{zz}-\gamma_c B)\tau)+ \sin^2(\theta_i/2) \right).
\end{equation}

This brings no benefits as compared to the simpler case achieved with a single spin by the Ramsey sequence because the nuclear spin itself is not a sensitive magnetometer. It is the post-selection that allows us to imprint the phase information into the nuclear spin, in such a way that variations of the post-selected angle $\theta_f$ calibrates the amount of information extracted from the measurement process~\cite{Coto17}. 

\section{Results}

The magnetic field sensitivity is defined as the minimum detectable 
magnetic field normalized by the total sequence time \cite{Degen17}. 
The minimum detectable magnetic field is found through the standard deviation $\Delta B$ \cite{Maze08,Aiello13},
\begin{equation}\label{variance}
\Delta B = \frac{\Delta I_z}{\vert\partial\langle I_z\rangle/\partial B \vert},
\end{equation}
where $\Delta I_z= \sqrt{\langle {I_z}^2\rangle - \langle I_z\rangle^2}$ is the standard deviation of the signal from the nuclear spin.

Clearly, in order to increase the sensitivity, we need a sharp response of the signal $\langle I_z\rangle$ to $B$ through the post-selection process. To begin with, we fix $B=10^{-2}$ G and for convenience, we choose both spins initially prepared in a superposition state with $\alpha=\theta_i=\pi/2$ and the post-selected state to be parallel to the pre-selected electronic spin state $\theta_f=\pi/2$. Without loss of generality, we consider a $^{13}$C nuclear spin with $A_{zz} = 500$~kHz. We can simplify the signal in Eq.~(\ref{exp_value}) with these parameters

\begin{equation}\label{exp_value2}
\langle I_z\rangle=	-\frac{\sin \left(\frac{A_{zz} \tau}{2}\right) \sin (\gamma_e B\tau)}{2 \left(\cos \left(\frac{A_{zz} \tau}{2}\right) \cos ( \gamma_e B\tau)+1\right)}.
\end{equation}

$A_{zz}$ contributes in two ways. Explicitly, it conduces to oscillations in the signal~\eqref{exp_value2}. Implicitly, it sets the upper limit of the nuclear spin control speed due to power broadening. As a result, stronger hyperfine coupling reduces the required nuclear spin gate time, and decreases the total sensing time of the protocol. 

\begin{figure}[ht]
\centering
\includegraphics[scale=0.4]{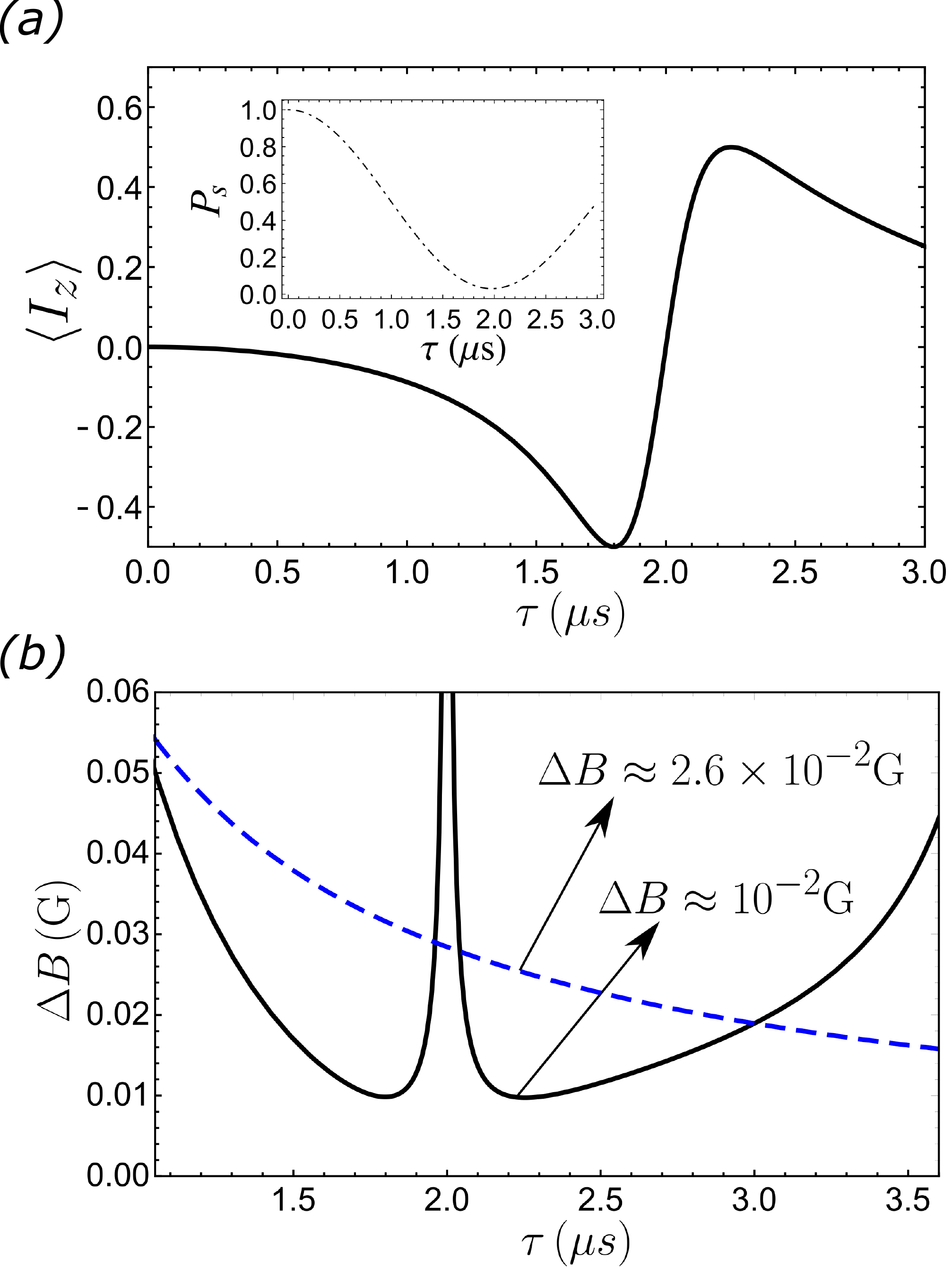}
\caption{ (a) Varying the interrogation time $\tau$ we are able to strongly modify the nuclear spin signal $\langle I_z \rangle$. Parameters are $\alpha=\theta_i=\theta_f=\pi/2$, $B=10^{-2}$ G. (b) The standard deviation of the magnetic field $\Delta B$ obtained from the post-selection (black-solid) has been improved as compared with the one obtained from Ramsey spectroscopy (blue-dashed), allowing high precision measurements. Values are taken at $\tau=2.2$ $\mu$s with a successful post-selection probability of $6\%$.}
\label{fig2}
\end{figure}

In Fig.~\ref{fig2} (a) we show the behavior of $\langle I_z\rangle$ as a function of the interrogation time $\tau$. Notice the region around $\tau=2$ $\mu$s where the signal is sharp. We numerically found that the optimal interrogation time is close to the extreme values of $\langle I_z\rangle$. The price to pay is a lower probability of successful post-selection ($P_s$). This trade-off between the signal gain and the probability of success is common in protocols that rely on post-selection, as stated in the field of weak value amplification \cite{Aharonov88,Combes14}. Consequently, many trials are required for the successful post-selection. Interestingly, even when outcomes are discarded, there is valuable information in the post-selection's statistics that can be used~\cite{Combes14,Alves15}, given that $P_s$ in Eq.~(\ref{norm}) is  a function of $B$. For more details see Appendix~\ref{Fisher}.

In Fig.~\ref{fig2} (b), we show the minimum detectable field allowed in our protocol obtained from Eq.~\eqref{variance}. One observes from the plot that our protocol constitutes a better route than the Ramsey sequence towards minimizing $\Delta B$ at short interrogation time ($\tau$), and this time is close to the coherence time of most NV centers. 
%It is important to notice that if $\tau$ can be increased, then Ramsey spectroscopy performs better. 
Therefore, concentrating valuable sensing information into a single successful measurement by post-selection provides a competitive approach for improving the minimum detectable magnetic field in a parameter range of interest.

The standard deviation $\Delta B$ does not take into account the number of failed post-selections, which increases the total time for the experiment. Hence, we now characterize our sensing protocol using the sensitivity ($\eta$) \cite{Maze08,Aiello13,Degen17}, 
\begin{equation}\label{sensitivity}
\eta=\Delta B \sqrt{t_m},
\end{equation}
where $t_m$ is the total sequence time consisting of initialization, interrogation and measurement, and includes those in failed post-selections. $t_m = N(t_i + \tau + t_p) + t_r$. The factor $N=1/P_s$ accounts for the average trials of the experiment for one successful post-selection. The initialization ($t_i$) and measurement times are usually fixed in a sequence, making the sensitivity dependent on the interrogation time. We split the measurement time for a single run into the post-selection time  $t_p$ (NV$^-$ readout) and nuclear spin readout time $t_r$. 

Firstly, we focus on the cryogenic temperature regime, $4$ K, where we can perform single-shot readout of the NV$^-$ electronic spin with high fidelity~\cite{Hensen15}. In combination with a nuclear spin controlled CNOT gate, one can readout the nuclear spin efficiently~\cite{Neumann10b,Maurer12,Dreau13}. We take $t_p=3.7\mu s$ for \NVm as reported in ref.~\cite{Hensen15}, and $t_r=5.7\mu s$ for $^{13}$C assuming a nuclear spin controlled CNOT gate limited by the hyperfine interaction strength in speed. 

The interrogation time $\tau$ is limited by transverse relaxation of the magnetometer, given by the characteristic time $T_2^\ast$ of the \NVm electronic spin. This relaxation process ($T_2^\ast$) for naturally occuring \NVm electronic spins in a natural abundance diamond sample is typically around a few microseconds~\cite{Blok14}. We model this pure dephasing process with the following Markovian master equation,

\begin{equation}\label{ME}
\frac{d\rho}{dt} = -i[\tilde{H},\rho] +  \Gamma(2S_z\rho S_z - S_z^2\rho -\rho S_z^2),
\end{equation} 
that introduces an exponential decay $\exp(-\Gamma t)$ on the off-diagonal elements of $\rho$, with $\Gamma=1/T_2^\ast$. More general non-Markovian magnetic noise can be modelled using a stochastic interaction ruled by the Ornstein-Uhlenbeck (OU) statistics \cite{deLange_science}. For instance, this type of noise has been observed in samples with high density of paramagnetic nitrogen centers (P1 centers) \cite{deLange_Scirep,Albrecht2014,Lei2017}. We discuss this case in Appendix~\ref{P1centers}.

For further comparison, once again we use the Ramsey sequence for reference, where the total sequence time reduces to $t_m=t_i+\tau+t_p$, with $t_i=1$ $\mu$s and $t_p=3.7$ $\mu$s.

\begin{figure}[ht]
\centering
\includegraphics[scale=0.55]{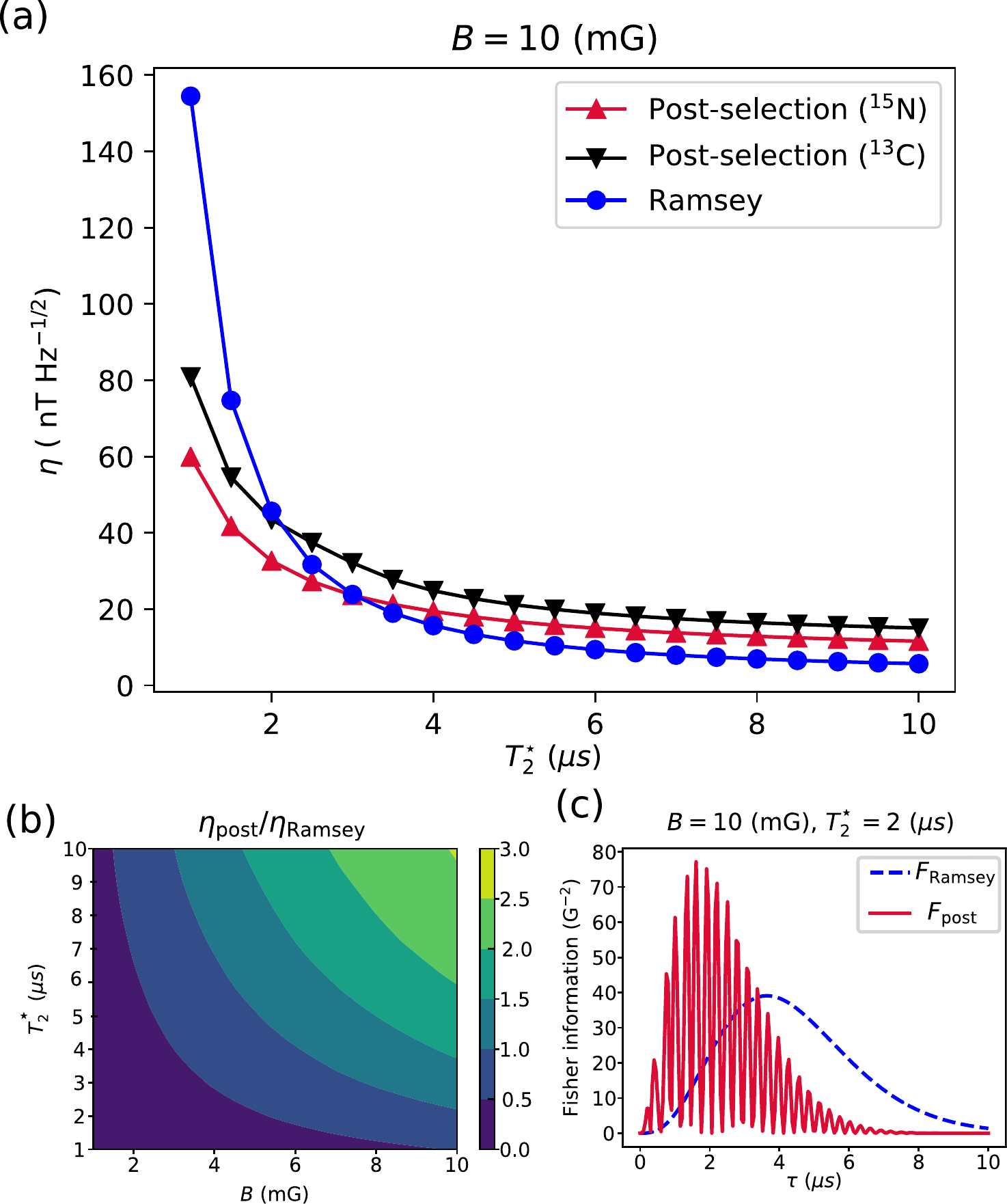}
\caption{ (a) Sensitivity as a function of the transverse relaxation $T_2^\ast$. The interrogation time $\tau$ is optimized for each point. Even when $T_2^\ast$ deteriorates the magnetometer, post-selection protocol remains comparable with Ramsey sequence. The total time for post-selection implemented with a \carb (\nitro) is $t_m=N(t_i+\tau+t_p)+t_r$, with $t_i=6$~$\mu$s ($t_i=1$~$\mu$s), $t_p= 3.7$~$\mu$s, $t_r= 5.7$~$\mu$s ($t_r=4.2$~$\mu$s) and $N=1/P_s(\tau)$. The total time for Ramsey is $t_m=t_i+\tau+t_p$, with $t_i=1$~$\mu$s and $t_p= 3.7$~$\mu$s. The other parameters are the same as in Fig.~\ref{fig2}. (b) The region where $\eta_{\rm post}/\eta_{\rm Ramsey}<1$ provides the parameter space for advantages in post-selection. (c) Fisher information shows that post-selection extract more information in shorter sensing time. }
\label{fig3}
\end{figure}

In Fig.~\ref{fig3} (a), we show the sensitivity for both Ramsey and post-selection protocols as a function of transverse relaxation time $T_2^\ast$. We consider a weak magnetic field $B =10^{-2}$ G and $\alpha=\theta_i=\theta_f=\pi/2$. Each value of sensitivity is taken at the optimal interrogation time ($\tau$). For long relaxation times $T_2^\ast$, the sensitivity decreases (improves), and Ramsey performs better. As $T_2^\ast$ decreases, the sensitivity of both magnetometers deteriorates, and for a suboptimal scenario $T_2^\ast\lesssim 2$ $\mu$s (for \carb) post-selection yields better sensitivity. It is worth noticing that Ramsey's optimal interrogation time is always greater than the one for post-selection.

So far we have considered a small magnetic field $B =10^{-2}$ G. In Appendix \ref{Appendix_Magn} we show the sensitivity as a function of a magnetic field up to $B=1$ G in the absence of losses. 

Hereafter, we consider the $T_2^\ast$ of naturally occuring \NVm electronic spins, $T_2^\ast = 2$ $\mu$s~\cite{Blok14}. In addition, we take into account the readout inefficiency by introducing a factor $C\leq 1$ and defining the sensitivity as $\eta_C=\eta/C$ \cite{Aiello13,Taylor08,Degen17}. Assuming ideal single-shot readout, this factor approaches to unity. Therefore, we can roughly estimate the sensitivity of our magnetometer for an optimal interrogation time $\tau=1.3$ $\mu$s to be $\eta=43.5$ nTHz$^{-1/2}$ ($C=1$). For a non-ideal scenario ($C=0.707$ \cite{Degen17}) we get $\eta_C=61.5$ nTHz$^{-1/2}$.

We remark that the protocol can be improved if we limit ourselves to cryogenic temperature applications only ($4$~K). We can replace the \carb meter by the native Nitrogen (assuming Nitrogen-$15$ (\nitro) here) \cite{Jacques09}. The \nitro~nuclear spin$-1/2$ is always present for the particular defect (\NVm), which allow us to make the protocol universal. Moreover, it only has isotropic hyperfine coupling $A_{zz}=3.03$ MHz, that is stronger than the \carb coupling ($A_{zz} = 500$~kHz), and thus enables faster nuclear spin gates. In Fig.~\ref{fig3} (a) we show the sensitivity using the \nitro. Due to the shorter initialization and gate times, the sensitivity with \nitro~is improved from that with \carb and it performs better than Ramsey in the suboptimal regime (short $T_2^\ast$). For further illustration, we calculate the ratio $\eta_{\rm post}/\eta_{\rm Ramsey}$ (for the \nitro) as a function of the relaxation time ($T_2^\ast$) and magnetic field ($B$), see Fig.~\ref{fig3} (b). One can observe now the region $\eta_{\rm post}/\eta_{\rm Ramsey}<1$, where our post-selection's based protocol exhibits an enhancement over the Ramsey sequence.

Next, we evaluate the performance of the protocol at room temperature. In this setting, we require the native Nitrogen nuclear spin as an ancilla and a high bias magnetic field ($B_z>2000$ G) to realize single shot readout for the post-selection through a repetitive readout protocol~\cite{Neumann10b}. With this protocol, single shot readout of the \NVm ~is achieved by firstly mapping the \NVm ~state to the Nitrogen and then read out the nuclear spin repetitively. Considering $2000$ repetitions we obtain $t_p= 5$ ms \cite{Neumann10b} and $t_r= 8$ ms \cite{Dreau13}, leading to $\eta_C=1.5$ $\mu$THz$^{-1/2}$. Based on numerical simulations we found the larger overhead for post-selection and readout does not change our conclusion fundamentally.

We further remark that the setting of our protocol, that is the NV center coupled to a nearby nuclear spin, is very versatile and has been explored extensively in both theories and experiments. For instance, the nuclear spin can provide different functionalities such as the quantum memory~\cite{Aslam17,Gurudev07}, the ancilla in quantum error correction~\cite{Taminiau14,Cramer16,Zhou18}, or the computational qubit itself~\cite{Bradley19}. In particular, recent works have taken advantages of the long coherence time of \carb nuclear spin and use it as a quantum memory to extend the interrogation time or refocus static noise~\cite{Matsuzaki16,Zaiser16}. 
All these functionalities, together with the post-selection protocol presented here, make the composite NV and nuclear spin system an ideal playground for exploring quantum information applications in a small scale. 

Finally, we introduce the Fisher information (FI) to quantify the amount of information that can be retrieved about the magnetic field $B$ for each protocol and show the results are consistent with the sensitivity analysis. FI can be calculated as \cite{Alves17}
\begin{equation}
    F=\sum_i{\frac{1}{P_i(B)}\left(\frac{\partial P_i(B)}{\partial B}\right)^2},
\end{equation}
where $P_i(B)$ is the probability of the measurement outcome $i$ and the sum is over all posible outcomes. For the case of post-selection, we multiply the FI by the probability of having a successful measurement ($P_s$ in Eq.~\eqref{norm}) \cite{Alves17}. In Fig.~\ref{fig3} (c) we show that the FI for post-selection is higher than the one for Ramsey, and maximum at short sensing times (same region where quantum Fisher information is maximum, see Appendix \ref{Fisher}). We can conclude that, in the presence of high decoherence, the post-selection protocol focuses more valuable information on a single successful measurement, which is available at a shorter sensing time, resulting in an advantage over the Ramsey protocol.

\section{Conclusions}

In summary, we have proposed a new experimentally feasible protocol based on post-selection to estimate a weak magnetic field. The information of the field is stored in the relative phase acquired by the electronic spin of a NV$^-$ center that is coupled to a nearby  nuclear spin. Using this protocol, the information regarding the magnetic field is focused on a single successful measurement upon the nuclear spin. Taking into account realistic conditions of losses and readout inefficiencies, we found that post-selection protocol is comparable with Ramsey in sensitivity in a wide range of transverse relaxation time $T_2^\ast$, and outperforms Ramsey for a range of $B<10^{-2}$ G and $T_2^\ast\lesssim 3$ $\mu$s. Furthermore, for this range our protocol results in a higher Fisher information that is achieved at shorter sensing times. We found that decreasing further initialization and gate times improves sensitivity, as shown in the case of native \nitro~nuclear spin. At cryogenic temperature ($4$ K), $B=10^{-2}$ G and $T_2^\ast=2$~$\mu$s the expected reduction (improvement) in sensitivity for post-selection with the \nitro~is around $28\%$ over Ramsey. At room temperature the most limiting factor is the number of repetitions for the readout of the nuclear spin. This could be improved by using Bayesian estimation \cite{Dinani19}. Moreover, nuclear spin introduces functionalities such as the quantum memory or the ancillae to implement error correction.

\section{acknowledgments}  RC acknowledges support from Fondecyt Iniciaci\'on No. 11180143. H.T.D. acknowledges support from Fondecyt-postdoctorado Grant No. 3170922. H.T.D and AN acknowledges support from Universidad Mayor through the Postdoctoral fellowship. JRM acknowledges support from Fondecyt Regular No. 1180673 and AFOSR FA9550-18-1-0513.

\appendix

\section{Multi-Rotating Frame}\label{Appendix_Rota}

The total Hamiltonian is given by
\begin{eqnarray}\label{eq:total_Ham}
H =& DS_z^2 + \gamma_e S_z(B_z + B) + \gamma_c I_z(B_z + B) + S_zA_{zz}I_z \nonumber\\
&+ \sqrt{2} \Omega^e S_x \cos(\omega_e t) + 2 \Omega_0^c I_x \cos(\omega_c t)\\
=& DS_z^2 + \gamma_e S_z(B_z + B) + \gamma_cI_z(B_z + B) + S_zA_{zz}I_z \nonumber\\
&+ \sqrt{2} \frac{\Omega^e}{2}S^+ e^{-i\omega_e t} +2 \frac{\Omega_0^c}{2}I^+ e^{-i\omega_c t} + h.c., 
\end{eqnarray}
where $\Omega^e$ and $\Omega_0^c$ are the Rabi frequencies of the electronic and nuclear spin transitions, respectively. $S^+=S_x+i S_y, I^+=I_x+i I_y$ and $S_i (I_i)$ are the electron (nuclear) spin-1 (-1/2) operators. In a rotating frame defined by the unitary operator $V=\exp[-i (\omega_e S_z + \omega_c I_z)t]$, the transformed Hamiltonian $\tilde{H}=V^\dagger H V +i(dV^\dagger/dt) V$ reads
\begin{eqnarray}
\tilde{H}=& (DS_z^2+D S_z) + \gamma_e B S_z + \gamma_c B I_z + S_zA_{zz}I_z \nonumber\\
&+ \sqrt{2} \frac{\Omega^e}{2}S_x +2 \frac{\Omega_0^c}{2}I_x .
\end{eqnarray}

Here we have taken $\omega_e=-D+\gamma_e B_z$ and $\omega_c=\gamma_c B_z$. After taking the rotating wave approximation, the transformed Hamiltonian can be explicitly written in a matrix form as

\begin{widetext}
\begin{equation}\label{eq:matrixform}
    \tilde{H}=\frac{1}{2}\begin{pmatrix}
    A_{zz} + 4D + (2\gamma_e+\gamma_c) B & 0 & 0 & 0 & 0 & 0\\
    0 & -A_{zz} + 4D + (2\gamma_e-\gamma_c) B & 0 & 0 & 0 & 0\\
    0 & 0 & \gamma_c B & \Omega_0^c & \Omega^e & 0\\
    0 & 0 & \Omega_0^c & -\gamma_c B & 0 & \Omega^e\\
    0 & 0 & \Omega^e & 0 & -A_{zz} + (\gamma_c -2\gamma_e) B & 0\\
    0 & 0 & 0 & \Omega^e & 0 & A_{zz} - (\gamma_c+2\gamma_e) B
    \end{pmatrix}.
\end{equation}
\end{widetext}

We now obtain the desired Hamiltonian in the $m_s=0,-1$ manifold by truncating this matrix. This leaves us
\begin{equation}\label{eq:eq3}
\tilde{H}=
\frac{1}{2}\begin{pmatrix}
\gamma_c B & \Omega_{0}^{c} & \Omega^{e}& 0  \\
 \Omega_{0}^c & -\gamma_c B & 0 & \Omega^{e}  \\
 \Omega^{e}& 0 & 2\delta_{1}^{\uparrow}& 0  \\
0 & \Omega^{e} & 0  & 2\delta_{1}^{\downarrow}
 \end{pmatrix},
\end{equation}

where $\delta_1^\uparrow=-\gamma_e B - A_{zz}/2 + \gamma_cB/2$ and $\delta_1^\downarrow=-\gamma_e B +A_{zz}/2 - \gamma_cB/2$.
%&\frac{1}{2}\begin{pmatrix}
%\gamma_c B & \Omega_0^c & \Omega^e & 0\\
%\Omega_0^c & -\gamma_c B & 0 & \Omega^e\\
%\Omega^e & 0 & -A_{zz} + (\gamma_c -2\gamma_e) B & 0\\
%0 & \Omega^e & 0 & A_{zz} - (\gamma_c+2\gamma_e) B
%\end{pmatrix}\\

\section{Fisher Information}\label{Fisher}

%We calculate the Fisher information for the postselection scheme and compare it with the case of Ramsey measurement. 
The Fisher information (FI) quantifies the amount of information that can be extracted about an unknown parameter from a measurement using a probe state. The quantum Fisher information (QFI) is the maximization of the FI over all the possible measurements \cite{Escher11}. According to the quantum Cram\'er-Rao bound \cite{Escher11,Braunstein} the variance in estimating a parameter, labeled by B here, using an unbiased estimative of B, is lower bounded by the inverse of the QFI, $var(B)\ge 1/\left(N F_Q\right)$, for $N$ measurements.

The FI is defined as 
\begin{equation}\label{eq:FI}
    F=\sum_i{\frac{1}{P_i(B)}\left(\frac{\partial P_i(B)}{\partial B}\right)^2},
\end{equation}
where $P_i(B)$ is the probability of the measurement outcome $i$ and the sum is over all the outcomes. The QFI is given by
\begin{equation}\label{eq:QFI}
    F_Q={\rm Tr}(\rho L^2),
\end{equation}
with $L$ the symmetric logarithmic derivative defined as 
\begin{equation}
    \frac{\partial \rho}{\partial B}=\frac{1}{2}(L\rho+\rho L).
\end{equation}
The matrix $L$ can be obtained by finding the eigenvalues $p_s$ and eigenvectors $|\psi_s\rangle$ of the density matrix, by which we can write  $\rho=\sum_s{p_s|\psi_s\rangle\langle\psi_s|}$. In this eigenbasis we have \cite{Liu}
\begin{equation}
    \langle\psi_i|\partial \rho|\psi_j\rangle =\frac{1}{2}(p_i+p_j)\langle\psi_i| L|\psi_j\rangle.
\end{equation}
The elements $L_{ij}=\langle\psi_i|L|\psi_j\rangle$ for which $p_i=p_j=0$ can be set to zero \cite{Liu14}.
For a pure state $|\psi\rangle$ the QFI can be simply written as \cite{Alves17,Liu}
\begin{equation}\label{eq:QFI_pure}
    F_Q=4\langle \partial_B\psi|\partial_B \psi\rangle-4|\langle\partial_B\psi|\psi\rangle|^2.
\end{equation}

For an electron spin of a \NVm center which is prepared in the state $|\psi\rangle=(|0\rangle+|1\rangle)/\sqrt{2}$ and evolves under the Hamiltonian $H=\gamma_e B S_z$ for time $\tau$, the QFI is $(\gamma_e \tau)^2$. In the absence of decoherence, the FI retrieved from a Ramsey sequence saturates this QFI, i.e., $F_{\rm Ramsey}=F_{Q}=(\gamma_e \tau)^2$, making Ramsey the optimal protocol. 

\begin{figure}[t!]
\includegraphics[scale=0.65]{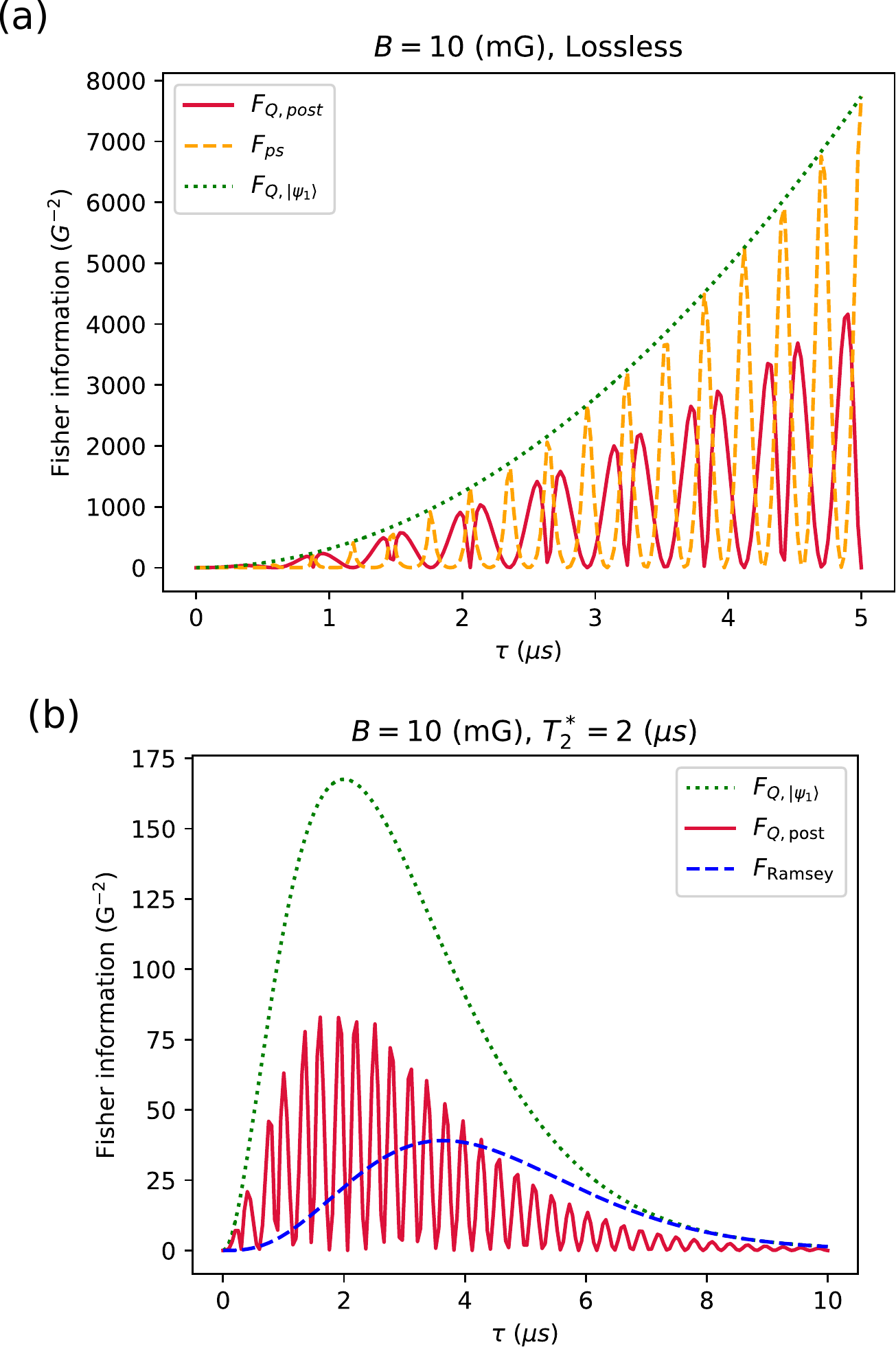}%{FI_plot_appendix.pdf}%{Figure6.eps}
\caption{Fisher information for $B=10$ mG and $\theta_f=\theta_i=\alpha=\pi/2$. (a) The QFI of the probe state $\ket{\Psi_1}$ (dotted green line) in Eq.~(5), the FI for the post-selection statistics (dashed orange line), and the QFI for the post-selected state (solid red line) in the absence of losses, $1/T^{\ast}_2\sim0$. (b) In the region where more information can be retrieved from $|\Psi_1\rangle$ (dotted green line), the QFI for the post-selection (solid red line) exhibits a significant enhancement over Ramsey's FI (dashed blue line), in the presence of losses  $T^{\ast}_2=2~\mu s$.}
\label{fig6}
\end{figure}

We now investigate the amount of information that can be extracted from the post-selection protocol with the \nitro. Firstly, we focus on the lossless condition, i.e. $1/T^\ast_2\sim0$ (in our numerical simulation we set $T^\ast_2=1$ s),  and we calculate the QFI for the state $\ket{\Psi_1}$ given in Eq.~(5) ($F_{{\rm Q}, \ket{\Psi_1}}$), which sets the upper bound. Then, we calculate the QFI resulting from the post-selection ($F_{{\rm Q, post}}$) (for the normalized post-selected state $|\phi_{post}\rangle$), that is weighted by the probability of success $P_s$ in Eq.~\eqref{norm}. In addition, we calculate the FI contained in the post-selection statistics ($F_{\rm ps}$). We show the results in Fig.~\ref{fig6} (a), and one can observe that there is a significant amount of information in the post-selection statistics $F_{\rm ps}$, but also in the favorable outcome $F_{{\rm Q, post}}$ at short sensing time. We remark that since the nuclear spin is not a good sensor itself, the QFI $F_{{\rm Q}, \ket{\Psi_1}}$ agrees with $F_{\rm Ramsey}$ in the absence of losses.

Secondly, we compare the FI for the post-selection protocol with the Ramsey sequence in the presence of a transverse relaxation process with $T^{\ast}_2=2$ $\mu$s, see Fig.~\ref{fig6} (b). It can be seen that in the region of time $\tau$ where more information can be retrieved, as given by the QFI $F_{{\rm Q}, \ket{\Psi_1}}$, post-selection ($F_{{\rm Q, post}}$) significantly outperforms Ramsey. We remark that the FI for post-selection is the same as its QFI $F_{{\rm Q, post}}$, as we have assumed that the post-selected state does not undergoes decoherence. The oscillations in $F_{{\rm Q, post}}$ originates from the large value of $A_{zz}$ of the \nitro. 

To conclude, these results confirm that even by only performing measurements on the post-selected state (and discarding the post-selection statistics) we can achieve an enhancement over the Ramsey sequence for ranges of $B$ and $T^{\ast}_2$ given by  $B<10$ mG and $T^{\ast}_2 \lesssim 3$ $\mu$s.

\section{Stochastic Noise}\label{P1centers}

\begin{figure}[ht]
	\centering
	\includegraphics[scale=0.30]{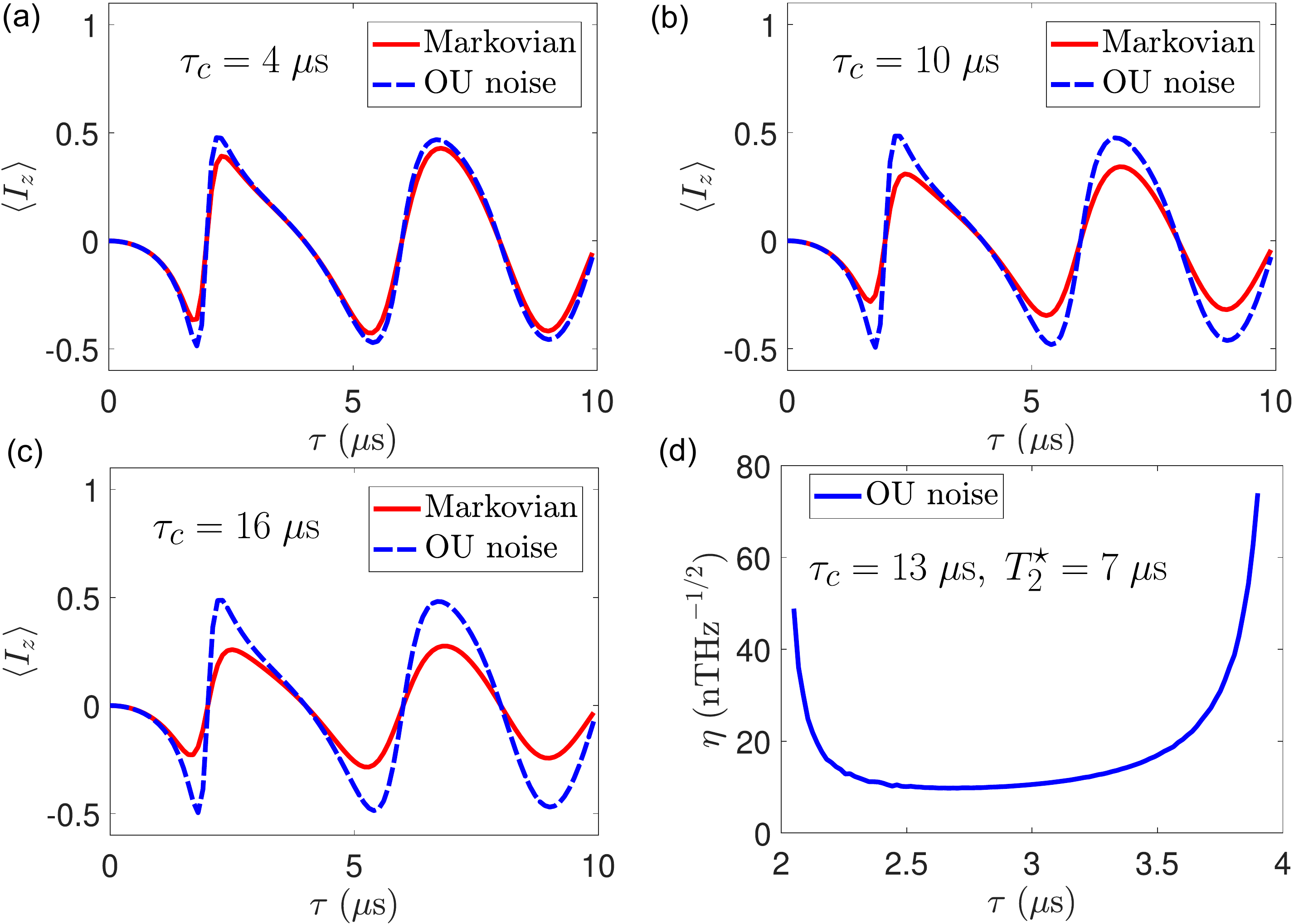}
	\caption{ (a)-(c) $\langle I_z\rangle$ for fixed $T_2^{\ast} = 20 \;\mu$s and $B=0.01$ G, for correlation times $\tau_c=4, 10, 16~\mu s$, respectively. The red and blue curves are the Markovian approximation and non-Markovian OU noise, respectively. (d) Sensitivity considering only Ornstein-Uhlenbeck noise in a highly non-Markovian regime.}
	\label{fig5}
\end{figure}

Non-markovian magnetic noise can be described using the following stochastic Hamiltonian ruled by the Ornstein-Uhlenbeck (OU) statistics ($\hbar = 1$)

\begin{eqnarray}
&H_{\rm noise}(t) = \gamma_e B(t) S_z,   \quad \langle B(t) \rangle = 0, \\ &\langle B(t)B(t') \rangle = B_s^2 e^{-|t-t'|/\tau_c} \nonumber,
\end{eqnarray}

where $\gamma_e$ is the electron gyromagnetic ratio, $B(t)$ is the stochastic magnetic field, $S_z$ is the spin operator for $S = 1$,  $B_s = \sqrt{\langle B(t)^2 \rangle}$ is the magnetic field intensity, and $\tau_c$ is the correlation time of the OU noise. It is clear that $H_{\rm noise}(t)$ commutes with the system Hamiltonian (\NVm and \carb) $H_s = \gamma_e B S_z + \gamma_n B I_z + S_z A_{zz} I_z$. Therefore, the time evolution of the system can be obtained through the application of the time evolution operator $U(t)=U_{\rm noise}(t)U_s(t)$. Here, $U_{\rm noise}(t) = \mbox{exp}\left(-i \int_{0}^{t} H_{\rm noise}(\tau) \, d\tau \right)$ and  $U_{\rm s}(t) = \mbox{exp}\left(-i H_{s} t \right)$. The magnetic noise can be generated using the recursive formula \cite{Albrecht2014,Gillespie}

\begin{equation}
B(t+dt) = B(t)e^{-dt/\tau_c} + \left[{c \tau_c \over 2}\left(1-e^{-2 dt/\tau_c} \right) \right]^{1/2}n,
\end{equation}

where $dt >0$ is the time step, $n$ is a normal random variable with mean value $0$ and variance $1$, and $c$ can be written in terms of the transverse relaxation time $T^*_2$ as $c = \frac{4}{(T_2^\ast)^2 \tau_c}$ \cite{deLange_science}. We note that the spin-bath interaction is intrinsically non-Markovian with a correlation time $\tau_c$. However, for evolution times $t\gg \tau_c$, the Markov approximation is valid, and the Lindblad super-operator associated with the stochastic Hamiltonian is given by
\begin{equation}
    \dot{\rho} = \mathcal{L}_{\rm markov}[\rho] = \gamma \left[S_z \rho S_z^{\dagger} - {1 \over 2}\{S_z S_z^{\dagger}, \rho\} \right],
\end{equation}
where $\mathcal{L}_{\rm markov}[\rho]$ is a pure-dephasing dissipation channel and $\gamma = 4  \tau_c/(T_2^\ast)^2$ is the dephasing rate. In Fig.~\ref{fig5} (a)-(c), we show the signal $\langle I_z \rangle$ in the presence of an OU noise using both a stochastic approach and Markovian approximation. We have used a range of correlation times for a system with $T_2^{\ast} = 20 \; \mu$s. We note that systems with $T_{2}^{\ast} \gg \tau_c$ are reasonably well described by the Markovian approximation, which is the case for the present work. Systems with a large memory time are beyond the scope of the proposed protocol. In such a case, a more complex envelope effect disturbs the signal $\langle I_z \rangle$. For instance, samples with high density of paramagnetic nitrogen centers, termed as P1 centers \cite{deLange_Scirep,Albrecht2014,Lei2017}, exhibits a correlation time $\tau_c=13$ $\mu$s and transverse relaxation time of few microseconds. In Fig. \ref{fig5} (d) we show the sensitivity as a function of the interrogation time ($\tau$) and observe a similar behavior as compared to the case modeled in the main text.  

\section{Varying Magnetic Field}\label{Appendix_Magn}

\begin{figure}[th]
	\centering
	\includegraphics[scale=0.45]{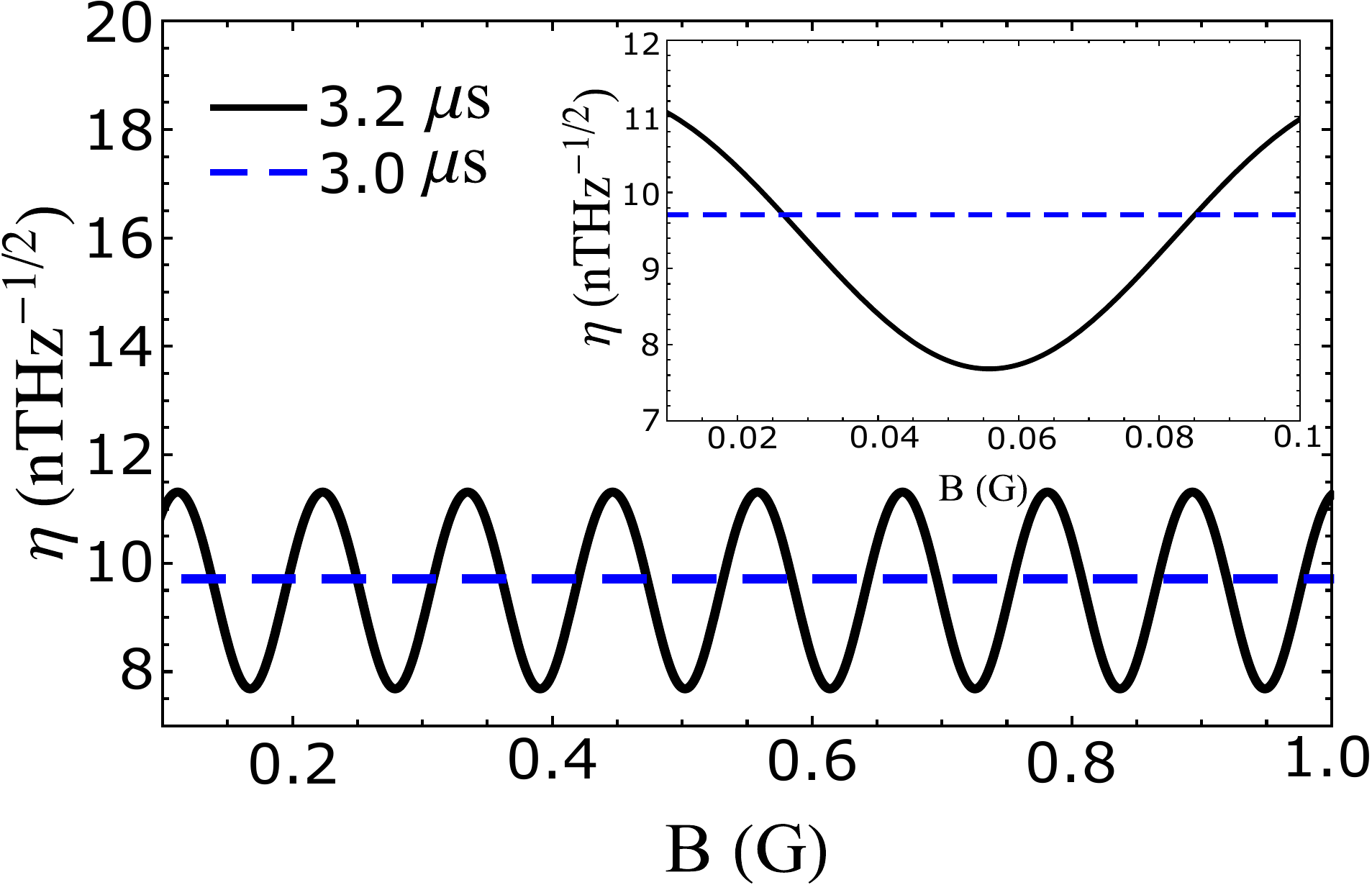}
	\caption{Magnetic field sensitivity for \carb in the range of $10^{-2}-10^{0}$ G, for $\tau=3.2$ and $\tau=3.0$ $\mu$s, and no losses.}
	\label{fig4}
\end{figure}

In this section we show the variation of the sensitivity ($\eta$) to detect the magnetic field as a function of the magnitude of the field, for fixed interrogation times. In Fig.~\ref{fig4}, the sensitivity remains constant ($\eta=9.7$ nTHz$^{-1/2}$) for $\tau=3.0$ $\mu$s, while for $\tau=3.2$ $\mu$s it exhibits oscillations. The magnetic field have been tuned in the range $10^{-2}-10^{0}$ G. Then, we can conclude that our protocol is suitable in a wide range of weak magnetic field sensing.

\bibliographystyle{unsrt}

\end{document}